\documentclass{article}
\usepackage{spconf,amsmath,graphicx}
\usepackage{multirow}
\usepackage{float}  
\usepackage{subfigure}  
\usepackage{epsfig}
\usepackage{verbatim}
\usepackage{bigstrut}
\usepackage{booktabs}

\title{PCF: ECAPA-TDNN with Progressive Channel Fusion for Speaker Verification}
\name{Zhenduo Zhao$^{1,2}$, Zhuo Li$^{1,2}$, Wenchao Wang${^1}$, Pengyuan Zhang$^{1,2}$\thanks{This work is partially supported by the National Key Research and Development Program of China(No. 2021YFC3320103)}}
\address{$^{1}$Key Laboratory of Speech Acoustics and Content Understanding, Institute of Acoustics, Chinese \\ Academy of Sciences, Beijing, China \\
  $^{2}$University of Chinese Academy of Sciences, Beijing, China}

\begin{document}
\maketitle
\begin{abstract}
ECAPA-TDNN is currently the most popular TDNN-series model for speaker verification, which refreshed the state-of-the-art(SOTA) performance of TDNN models. However, one-dimensional convolution has a global receptive field over the feature channel. It destroys the time-frequency relevance of the spectrogram. Besides, as ECAPA-TDNN only has five layers, a much shallower structure compared to ResNet restricts the capability to generate deep representations. To further improve ECAPA-TDNN, we propose a progressive channel fusion strategy that splits the spectrogram across the feature channel and gradually expands the receptive field through the network. Secondly, we enlarge the model by extending the depth and adding branches. Our proposed model achieves EER with 0.718 and minDCF(0.01) with 0.0858 on vox1o, relatively improved 16.1\% and 19.5\% compared with ECAPA-TDNN-large.
\end{abstract}
\section{Introduction}
\label{sec:intro}

Speaker verification aims to determine whether a piece of speech belongs to the claimed speaker. As an important method of biometric authentication, it has a broad wide range of application scenarios. 

Automatic speaker verification(ASV) systems consist of three modules, an embedding extractor, a scoring backend, and a calibration module. In recent years, ASV systems based on deep neural networks have continuously refreshed the SOTA performance. To achieve better performance, researchers have innovated on each module to push up the upper limit of performance. The embedding extractor is the key component of a system and contributes the most among all modules. Starting from x-vector\cite{xvector}, plenty of works have been proposed on building more powerful networks. These architectures can be roughly divided into one-dimensional convolution networks, two-dimensional convolution networks, and attention-based transformer\cite{transformer}. While transformer gives a less competitive performance on ASV tasks without large-scale pre-trained models, convolution-based structures take the mainstream position. There are diverse aspects to boost system performance, adding more layers \cite{etdnn,ftdnn,nectt_sre19,dfresnet} helps model extracting deep representations, adding residual connections \cite{resnet} make it faster for convergence while avoiding gradient vanishing, introducing attention module \cite{se,hsnet,ecapa,fbatt} improves the ability to capture long-range dependencies and so on. 

ECAPA-TDNN\cite{ecapa} improves performance by introducing several useful methods. However, compared with ResNet models, we find out that it could be further improved. Firstly, we argue that the limitation of one-dimensional convolution restricts its performance. Compared with two-dimensional convolution, TDNN has a global sensor space over the feature channel, and thus lost time-frequency correlation at the first block. Secondly, ECAPA-TDNN only has five blocks, which restricts the generation of deep representations. Finally, the use of different sizes of convolution kernel in the same layer \cite{repvgg} could improve the ability to capture multi-scale features. 

This paper proposes a simple but effective strategy called progressive channel fusion(PCF). This strategy splits the input spectrogram into several frequency bands and progressively fuses the bands through the network. Thereby it gets a local receptive field over both time and frequency channels similar to ResNet and reduces the parameter number at the same time. Besides, we introduce two useful methods that further improve the performance accompanied by PCF, res2block branching, and block deepening.

This paper is organized as follows: Section 2 describes related works, Section 3 describes the proposed method, Section 4 presents experiments and results, and Section 5 summarizes and looks to future work.

\begin{figure}
    \centering
    \includegraphics[width=6.6cm]{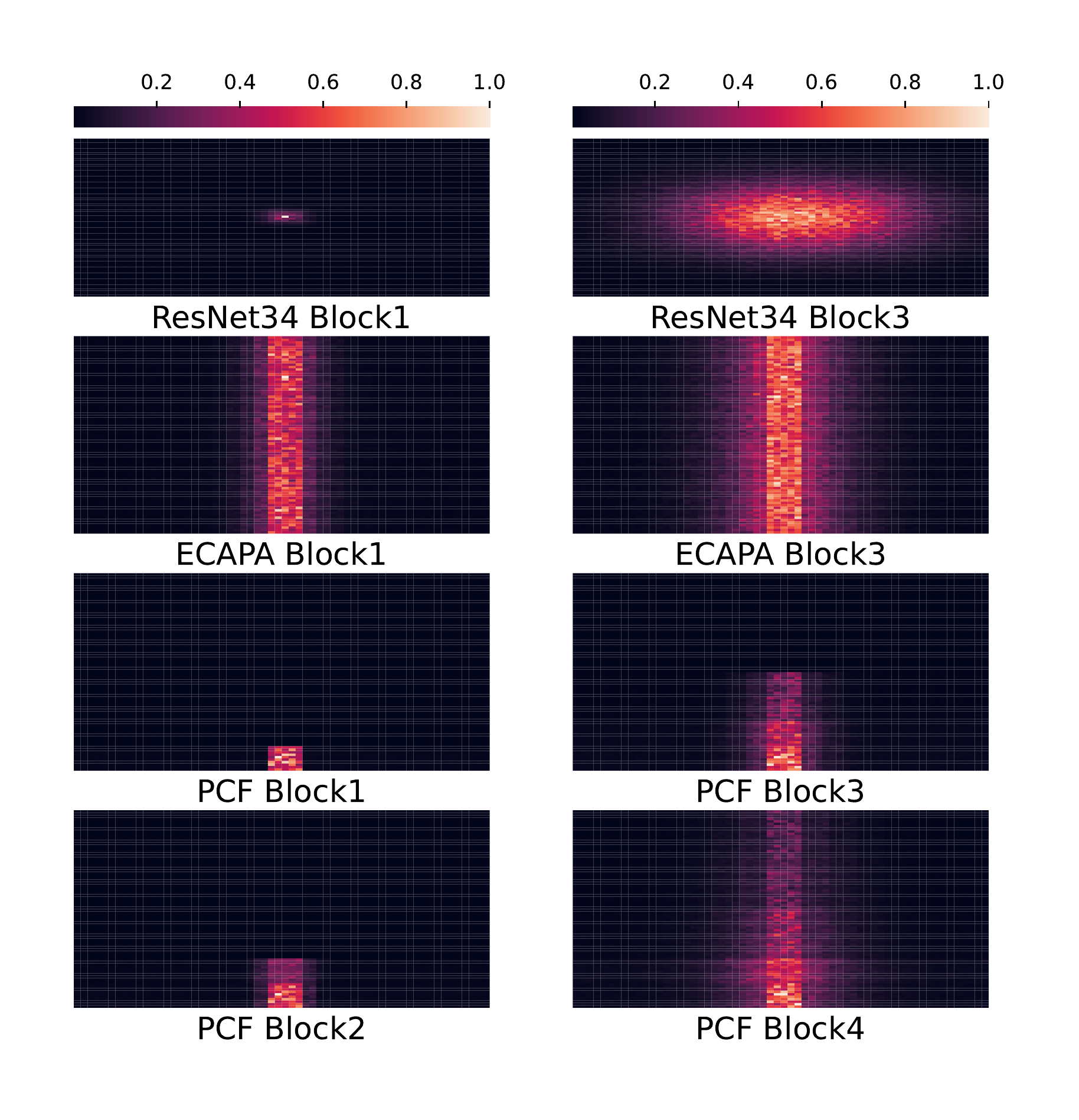}
    \caption{The receptive field of ResNet34 with 64 base channels, ECAPA-TDNN (C=1024), and our PFA-ECAPA (C=1024). For the first two models, we only give the results from blocks 1\&3, and for PFA-ECAPA, we give all four layers' sensor space.}
    \label{fig:my_label}
\end{figure}

\section{Related Work}
\label{sec:related}

We adopt two architectures as the baseline: ECAPA-TDNN and ResNet.

ECAPA-TDNN introduces multiple module construction strategies. We summarize the key points as follows:
\begin{itemize}
    \item Channel- and context-dependent statistics pooling. It extends the attention mechanism to channel dimension, to make the model focus more on speaker characteristics instead of non-active frames. Besides, it introduces global context representation by concatenating local input and global non-weighted mean and standard deviation of the feature map across the time domain, thereby taking global properties into account, such as background noises or recording conditions.

    \item Res2Net\cite{gao2019res2net} blocks. By splitting input channels into several pieces, and hierarchically executing convolution, addition, and concatenation, the Res2Block module enhances the capability to capture multi-scale features while significantly reducing the number of parameters.
    
    \item Squeeze-Excitation(SE) \cite{se} module. SE module first generates a channel-wise descriptor, called the squeeze operation. Then, it computes channel weights based on the descriptor, and applied them to the original channels, called the excitation operator. 
    
    \item Multi-layer feature aggregation(MFA) module. Shallow layer outputs in deep neural networks also contain speaker-relevant information. MFA module concatenates outputs from 3 SE-Res2Net blocks for robustness improvement.
\end{itemize}

ResNet models are dominant in recent speaker recognition challenges. Benefiting from residual connection and modular design, ResNet can easily scale to large while maintaining its ability of fast convergence. After its backbone, we use the same attentive statistics pooling as ECAPA-TDNN for comparison.

\section{Proposed Method}
\label{sec:proposed}

We propose a progressive channel fusion strategy, branch res2block, and layer deepening methods to further enhance the performance of ECAPA-TDNN. The topology of the model is shown in Fig 2 and the relevant configuration gives in Table 1.

\begin{figure}
  \centering
  \includegraphics[height=10cm]{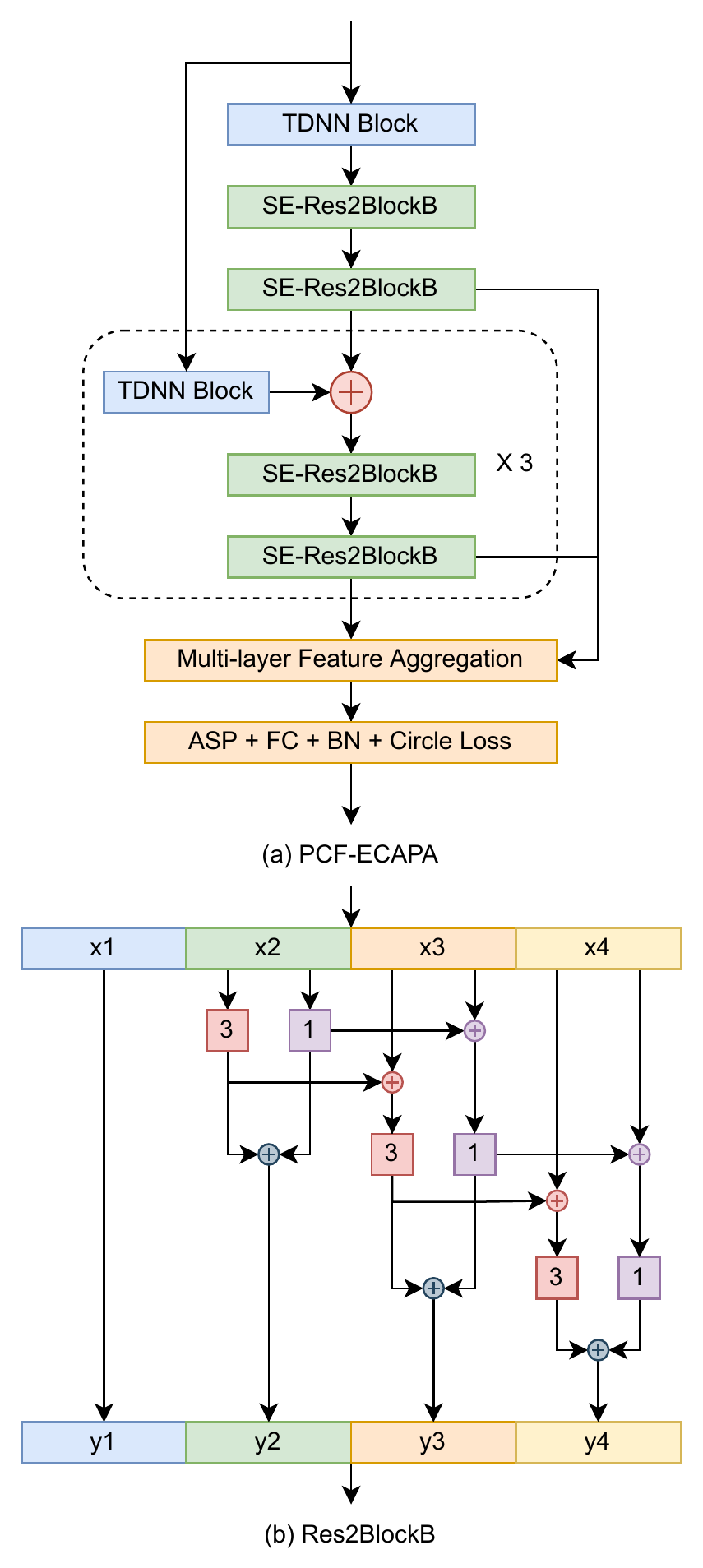}
  \caption{Proposed Structure. TDNN block is a sequential of a conv1d, a ReLU and a batchnorm1d. Res2Block in ECAPA-TDNN is replaced with Res2BlockB shown in (b)}
\label{fig:model arch}
\end{figure}

\begin{table}[htb]
    \centering
    \begin{tabular}{cc}
    \toprule
    Layer & Structure \\
    \midrule
    Input & $B \times F \times T$ \\
    \midrule
    Link1 & TDNN-Block(80,C,5,1,8) \\
    Link2 & TDNN-Block(80,C,5,1,4) \\
    Link3 & TDNN-Block(80,C,5,1,2) \\
    Link4 & TDNN-Block(80,C,5,1,1) \\
    \midrule
    Block1 & SE-Res2BlockB(C,C,3,1,8)$\times$2 \\
    Block2 & SE-Res2BlockB(C,C,3,2,4)$\times$2 \\
    Block3 & SE-Res2BlockB(C,C,3,3,2)$\times$2 \\
    Block4 & SE-Res2BlockB(C,C,3,4,1)$\times$2 \\
    MFA & TDNN-Block(4$\times$C,C,1,1,1) \\
    Pooling & ASP(C,1536) \\
    \midrule
    FC & Conv1D(1536,192,1,1,1) \\
    FC & Linear(192,N)\\
    \bottomrule
    \end{tabular}
    \caption{Proposed model structure. The symbols in brackets represent input channels, output channels, kernel size, dilation, number of sub-bands for Convolution blocks, and input dim and output dim for linear blocks.N means speaker number.}
    \label{tab:my_label}
\end{table}


\subsection{Progressive Channel Fusion}
The local receptive field is one of the basis when constructing deep convolution networks, in collaboration with stacked blocks, it expands sensor space from local to the whole feature map in deep blocks while avoiding over-fitting. One of the critical differences between ResNet and TDNN is that two-dimensional convolution preserves a local receptive field on both frequency and temporal channels. Because conventional TDNN has full access to all feature channels, the risk of over-fitting goes high when increasing the number of parameters. To alleviate this phenomenon, we propose a strategy called progressive channel fusion, which allows the model to gradually fuse the channel relationship across the forward propagation. 

Assume input feature map $X \in \mathcal{R}^{F \times T}$, where $F, T$ is the feature dimension and temporal dimension respectively. We split $F$ into $N$ sub-bands, where $N=8$ for the first block, then we half $N$ after each block. It should be noted that in Res2Block, benefit from its hierarchical design, frequency bands still have the opportunity to communicate, while the difference is that our strategy assigns major frequency bands to channel splits. Besides, we introduce links between the spectrogram and each block with a TDNN block, it shares an identical split configuration with the target block. The receptive field of ResNet34, ECAPA-TDNN, and the proposed PCF-ECAPA are visualized in Fig 1. the first row shows the receptive field of block1 and block3 in ResNet34. It has a gradually increased sensor space. The second row gives the result of the original ECAPA-TDNN, which always has global sensor space across the channel dimension, while the last two rows present all blocks' receptive field of our proposed progressive channel fusion strategy. It shares a similar behavior with the ResNet model, where the visible area of the spectrogram spread around as the block goes deep.

\subsection{Res2Block Branching} 
RepVGG \cite{repvgg,zhao2021speakin} has proven to be effective in recent challenges. Due to its re-parameterized structure design, the model can learn multi-scale features during training, and the convolution branches are merged into a single branch for inference. The key operation to improve the performance of RepVGG is the multi-branch structure, it introduces convolution branches with multiple kernel sizes so that the model can learn multi-scale representation. Therefore, we also introduce this structure by adding a branch with a convolution kernel size of 1 in the form of res2block. Although the two branches cannot be merged into one during inference, it still boosts the ability to capture input features at multiple levels.

\subsection{Layer Deepening} 
It is usually more effective to make the model deeper rather than wider when scaling convolution networks because of the growing range of receptive field. In section 3.1, we introduce a progressive channel fusion strategy, which improves the model performance while reducing the model parameters. To make up for the weakening of the modeling ability caused by the reduction of parameters, after the first TDNN block, we extend the backbone to 4 blocks, each block containing 2 Res2Blocks with the same dilation. For the MFA module, we concatenate outputs from 4 blocks as the input.

\section{Experiments}
\label{sec:expe}

\begin{table*}[htb]
\caption{PCF-ECAPA Performance on VoxCeleb Official Evaluation Sets}
\label{tab:res2net50result}
\centering
\small
\setlength{\tabcolsep}{1.2mm}{
\begin{tabular*}{\linewidth}{lccccccccccccccc}
  \toprule
  \multirow{2}*{Stage} & \multirow{2}*{Params} & \multicolumn{2}{c}{Vox1O} & \multicolumn{2}{c}{Vox1E} & \multicolumn{2}{c}{Vox1H} & \multicolumn{2}{c}{Vox20-dev} & \multicolumn{2}{c}{Vox21-dev} & \multicolumn{2}{c}{Vox22-dev} \\
  \cmidrule{3-15}
  & & EER & DCF$_{0.01}$ & EER & DCF$_{0.01}$ & EER & DCF$_{0.01}$ & EER & DCF$_{0.01}$ & EER & DCF$_{0.01}$ & EER & DCF$_{0.01}$\\
  \midrule
  ResNet18 & 6.7M & 1.510 & 0.1789 & 1.559 & 0.1760 & 2.679 & 0.2642 & 4.235 & 0.3540 & 4.835 & 0.3798 & 3.624 & 0.4142  \\
  ResNet34 & 9.3M & 1.164 & 0.1141 & 1.167 & 0.1285 & 2.099 & 0.2127 & 3.365 & 0.2794 & 3.806 & 0.3050 & 2.825 & 0.3062 \\
  \midrule
  ECAPA(C=512) & 6.2M & 1.058 & 0.1021 & 1.205 & 0.1371 & 2.182 & 0.2155 & 3.537 & 0.2905 & 4.545 & 0.3643 & 2.979 & 0.3541 \\
  ECAPA(C=1024) & 14.7M & 0.856 & 0.1066 & 1.074 & 0.1285 & 2.009 & 0.2021 & 3.265 & 0.2725 & 4.142 & 0.3353 & 2.830 & 0.3110 \\
  \midrule
  PCF-ECAPA(C=512) & 8.9M & \textbf{0.718} & \textbf{0.0858} & \textbf{0.792} & 0.1138 & 1.802 & \textbf{0.1750} & 2.959 & \textbf{0.2250} & 3.684 & 0.3073 & 2.630 & 0.2836 \\
  PCF-ECAPA(C=1024) & 22.2M & \textbf{0.718} & 0.0892 & 0.891 & \textbf{0.1024} & \textbf{1.707} & 0.1754 & \textbf{2.831} & 0.2339 & \textbf{3.527} & \textbf{0.2880} & \textbf{2.333} & \textbf{0.2666} \\
  \bottomrule
\end{tabular*}}
\end{table*}

\subsection{Dataset}
We use voxceleb2-dev \cite{voxceleb2} as training set, containing 1,092,009 utterances from 5,994 speakers. For data augmentation, we use MUSAN \cite{musan} and RIRS-NOISES \cite{rirs}. 80-dimensional log-Mel Filter Bank is used as input features with cepstral mean normalization applied, and voice activity detection is not used. For evaluation, we use official evaluation sets including Vox1o,e,h, and validation sets from the last three years of VoxCeleb speaker recognition challenges\cite{voxsrc2020,voxsrc2021}.

\subsection{Model}
We use the ECAPA-TDNN implemented by speechbrain \cite{speechbrain} as the baseline model for the experiments, where two settings are used, a base model with 512 channels, and a large model with 1024 channels. Our model also adopts the same configuration. We fix the output channels of the MFA module to 1536 to be consistent with the original configuration. For ResNet, we use ResNet18 and ResNet34 with 32 channels and the same pooling as ECAPA-TDNN.

\subsection{Training}
We use Adam \cite{adam} for optimization, and the learning rate curve adopt the cycle strategy \cite{cycle}. We train for 3 cycles with one cycle lasting 100k steps, the learning rate in each cycle varies from 1e-8 to 1e-3, and the weight decay is 5e-5. The batch size is set to 256. Circle loss \cite{circle}, which has stronger constraints on speaker embedding compared to AAM-Softmax loss \cite{aam}, is used with $m=0.35,s=60$. All models are trained with the same setting.

\subsection{Evaluation}
To bridge the gap between the duration of segments for training and evaluation, we sample short clips from testing utterances and use the mean of the cosine similarity between a pair of embedding matrices as the final score. We test the model at the end of the final cycle and report all systems performance in terms of equal error rate(EER) and minimum Detection Cost Function(minDCF) with $p_{target}=0.01, C_{FA} = C_{Miss} = 1$.

\subsection{Results}

Table 2 summarizes the performance of PCF-ECAPA and the original ECAPA-TDNN together with the number of model parameters except for the classification layer. Our proposed architecture outperforms the baseline systems and gives an average relative improvement of 15.6\% on EER and 15.2\% on minDCF over the best baseline system. 
We conducted ablation experiments on the base model, and the results are shown in Table 3. From the ECAPA-TDNN base, we stack three methods one by one and finally get the proposed PCF-ECAPA.

We first evaluate the impact of the model depth. Simply increasing the number of blocks from 3 to 8 dramatically improves the performance, where EER improves from 1.058 to 0.792 and minDCF reduces to 0.0912, giving 26.6\% and 10.7\% relative improvement respectively. The almost doubled number of parameters greatly enhances the model's ability to capture deep representations. Meanwhile, the deepened base model exceeds ECAPA-TDNN-large for 7.5\% with less amount of parameters. It proves our assumption that it is usually more efficient for convolution networks to deepen the model rather than broaden it. 

Secondly, the branched block pushes the EER to 0.744 but pulls the minDCF back to 0.1024. Parallel branch structure helps the model capture multi-scale features at the cost of 1.8\% extra parameters. The performance loss may come from the circle loss based on our experience.

Finally, we apply the proposed PCF strategy. It further improves EER to 0.718 and minDCF to 0.0858. Restricting the sensor space of TDNN models over channel dimension is proved to be effective because of the fine-grained attention on each frequency sub-bands. Besides, it brings an 18.3\% cut on the number of parameters as a result of the local receptive field.

Moreover, experiments show that simply scaling the model from 512 channels to 1024 channels brings little improvement,  resulting in EER=0.718 and minDCF=0.0892 on vox1o. On other evaluation sets, the large model gets the biggest boost on Vox22-dev. Nevertheless, it has a less competitive parameter efficiency.

\begin{table}[htbp]
  \centering
  \small
  \caption{Ablation Study of the PCF-ECAPA architecture. A represents deepen model, B represents adding branches, and C represents using PCF strategy. With ABC applied, we get PCF-ECAPA}
    \begin{tabular}{lccc}
    \toprule
    Systems & Params & EER & minDCF(0.01) \\
    \midrule
    Base & 6.2M & 1.058 & 0.1021 \\
    \midrule
    +A & 10.7M & 0.792 & 0.0912 \\
    ++B & 10.9M & 0.744 & 0.1024 \\
    +++C & 8.9M & 0.718 & 0.0858 \\
    \bottomrule
    \end{tabular}%
  \label{t-base}%
\end{table}%

\section{Conclusion}
\label{sec:conclusion}

In summary, we propose a strategy to enhance TDNN models: progressive channel fusion. This method enables the model to pay attention to the narrow frequency band in shallow layers, gradually expand the receptive filed through the network, and have the global frequency band receptive field in deep layers, thereby improving the overall utilization efficiency of the feature map by the model. In addition, we introduce the branch structure and deepen the number of model layers to further improve the model performance, and refreshed the SOTA of TDNN models with all three methods stacked. Experiments show that our optimization direction of the model structure is correct and still has the potential for better performance.

\vfill\pagebreak

\bibliographystyle{IEEEbib}
\bibliography{main}

\begin{thebibliography}{10}

\bibitem{xvector}
David Snyder, Daniel Garcia-Romero, and Gregory et~al Sell,
\newblock ``X-vectors: Robust dnn embeddings for speaker recognition,''
\newblock in {\em 2018 IEEE international conference on acoustics, speech and
  signal processing (ICASSP)}. IEEE, 2018, pp. 5329--5333.

\bibitem{transformer}
Ashish Vaswani, Noam Shazeer, and Niki et~al Parmar,
\newblock ``Attention is all you need,''
\newblock {\em Advances in neural information processing systems}, vol. 30,
  2017.

\bibitem{etdnn}
David Snyder, Daniel Garcia-Romero, Gregory Sell, Alan McCree, Daniel Povey,
  and Sanjeev Khudanpur,
\newblock ``Speaker recognition for multi-speaker conversations using
  x-vectors,''
\newblock in {\em ICASSP 2019-2019 IEEE International conference on acoustics,
  speech and signal processing (ICASSP)}. IEEE, 2019, pp. 5796--5800.

\bibitem{ftdnn}
Daniel Povey, Gaofeng Cheng, and Yiming et~al Wang,
\newblock ``Semi-orthogonal low-rank matrix factorization for deep neural
  networks.,''
\newblock in {\em Interspeech}, 2018, pp. 3743--3747.

\bibitem{nectt_sre19}
Kong~Aik Lee, Koji Okabe, and Hitoshi et~al Yamamoto,
\newblock ``Nec-tt speaker verification system for sre'19 cts challenge.,''
\newblock in {\em INTERSPEECH}, 2020, pp. 2227--2231.

\bibitem{dfresnet}
Bei Liu, Zhengyang Chen, and Shuai et~al Wang,
\newblock ``Df-resnet: Boosting speaker verification performance with
  depth-first design,''
\newblock {\em Inter-Speech 2022}, 2022.

\bibitem{resnet}
Kaiming He, Xiangyu Zhang, Shaoqing Ren, and Jian Sun,
\newblock ``Deep residual learning for image recognition,''
\newblock in {\em Proceedings of the IEEE conference on computer vision and
  pattern recognition}, 2016, pp. 770--778.

\bibitem{se}
Jie Hu, Li~Shen, Samuel Albanie, Gang Sun, and Enhua Wu,
\newblock ``Squeeze-and-excitation networks,''
\newblock {\em computer vision and pattern recognition}, 2018.

\bibitem{hsnet}
Zhuo Li,
\newblock ``Explore long-range context feature for speaker verification,''
\newblock {\em CoRR}, vol. abs/2112.07134, 2021.

\bibitem{ecapa}
Brecht Desplanques, Jenthe Thienpondt, and Kris Demuynck,
\newblock ``Ecapa-tdnn: Emphasized channel attention, propagation and
  aggregation in tdnn based speaker verification,''
\newblock {\em Proc. Interspeech 2020}, pp. 3830--3834, 2020.

\bibitem{fbatt}
Aiwen Deng, Shuai Wang, Wenxiong Kang, and Feiqi Deng,
\newblock ``On the importance of different frequency bins for speaker
  verification,''
\newblock in {\em ICASSP 2022-2022 IEEE International Conference on Acoustics,
  Speech and Signal Processing (ICASSP)}. IEEE, 2022, pp. 7537--7541.

\bibitem{repvgg}
Xiaohan Ding, Xiangyu Zhang, and Ningning et~al Ma,
\newblock ``Repvgg: Making vgg-style convnets great again,''
\newblock in {\em Proceedings of the IEEE/CVF Conference on Computer Vision and
  Pattern Recognition}, 2021, pp. 13733--13742.

\bibitem{gao2019res2net}
Shang-Hua Gao, Ming-Ming Cheng, Kai Zhao, Xin-Yu Zhang, Ming-Hsuan Yang, and
  Philip Torr,
\newblock ``Res2net: A new multi-scale backbone architecture,''
\newblock {\em IEEE transactions on pattern analysis and machine intelligence},
  vol. 43, no. 2, pp. 652--662, 2019.

\bibitem{zhao2021speakin}
Miao Zhao, Yufeng Ma, Min Liu, and Minqiang Xu,
\newblock ``The speakin system for voxceleb speaker recognition challange
  2021,''
\newblock {\em arXiv preprint arXiv:2109.01989}, 2021.

\bibitem{voxceleb2}
Joon~Son Chung, Arsha Nagrani, and Andrew Zisserman,
\newblock ``Voxceleb2: Deep speaker recognition,''
\newblock {\em conference of the international speech communication
  association}, 2018.

\bibitem{musan}
David Snyder, Guoguo Chen, and Daniel Povey,
\newblock ``Musan: A music, speech, and noise corpus,''
\newblock {\em arXiv: Sound}, 2015.

\bibitem{rirs}
Tom Ko, Vijayaditya Peddinti, Daniel Povey, Michael~L. Seltzer, and Sanjeev
  Khudanpur,
\newblock ``A study on data augmentation of reverberant speech for robust
  speech recognition,''
\newblock {\em international conference on acoustics, speech, and signal
  processing}, 2017.

\bibitem{voxsrc2020}
Arsha Nagrani, Joon~Son Chung, Jaesung Huh, Andrew Brown, Ernesto Coto, Weidi
  Xie, Mitchell McLaren, Douglas~A Reynolds, and Andrew Zisserman,
\newblock ``Voxsrc 2020: The second voxceleb speaker recognition challenge,''
\newblock {\em arXiv preprint arXiv:2012.06867}, 2020.

\bibitem{voxsrc2021}
Andrew Brown, Jaesung Huh, Joon~Son Chung, Arsha Nagrani, and Andrew Zisserman,
\newblock ``Voxsrc 2021: The third voxceleb speaker recognition challenge,''
\newblock {\em arXiv preprint arXiv:2201.04583}, 2022.

\bibitem{speechbrain}
Mirco Ravanelli, Titouan Parcollet, and Peter~Plantinga et~al,
\newblock ``{SpeechBrain}: A general-purpose speech toolkit,'' 2021,
\newblock arXiv:2106.04624.

\bibitem{adam}
Diederik~P. Kingma and Jimmy Ba,
\newblock ``Adam: A method for stochastic optimization,''
\newblock {\em arXiv: Learning}, 2014.

\bibitem{cycle}
Leslie~N. Smith,
\newblock ``Cyclical learning rates for training neural networks,''
\newblock {\em workshop on applications of computer vision}, 2015.

\bibitem{circle}
Runqiu Xiao, Xiaoxiao Miao, and Wenchao~Wang et~al,
\newblock ``{Adaptive Margin Circle Loss for Speaker Verification},''
\newblock in {\em Proc. Interspeech 2021}, 2021, pp. 4618--4622.

\bibitem{aam}
Jiankang Deng, Jia Guo, Niannan Xue, and Stefanos Zafeiriou,
\newblock ``Arcface: Additive angular margin loss for deep face recognition,''
\newblock {\em arXiv: Computer Vision and Pattern Recognition}, 2018.

\end{thebibliography}

\end{document}